\documentclass[prl,twocolumn,aps,epsfig,prbbib]{revtex4}
\usepackage{graphicx}

\begin{document}

\title{Magnetically Mediated Transparent Conductors: \\
In$_2$O$_3$ doped with Mo}

\author{J.~E. Medvedeva}\email[]{E-mail:juliaem@umr.edu}

\affiliation{Department of Physics, University of Missouri--Rolla, Rolla, MO 65409}


\begin{abstract}
First-principles band structure investigations of the electronic, 
optical and magnetic properties of Mo-doped In$_2$O$_3$ reveal the vital role of magnetic 
interactions in determining both the electrical conductivity and the Burstein-Moss shift 
which governs optical absorption. We demonstrate the advantages of the transition metal 
doping which results in smaller effective mass, 
larger fundamental band gap and better overall optical transmission in the visible 
-- as compared to commercial Sn-doped In$_2$O$_3$. Similar behavior is expected upon doping 
with other transition metals opening up an avenue for the family of efficient transparent 
conductors mediated by magnetic interactions.

{\bf PACS number(s)}: 71.20.-b  

\end{abstract}

\maketitle

A unique combination of high electrical conductivity and low optical absorption 
-- a property which makes transparent conducting oxides (TCO) attractive 
for many advanced applications 
\cite{Thomas,MRS,Stoute} -- is well-known to be 
a challenge due to the mutual dependence of the optical transmission rates 
and conductivity.
Among possible routes to avoid compromising the optical transparency
\cite{Coutts,nano} is to enhance conductivity via 
mobility of the free carriers rather than their concentration.
Recently, the mobility 
with more than twice the value of the typical commercial TCO materials such as
Sn-doped indium oxide (ITO) was observed in Mo-doped In$_2$O$_3$ (IMO),
and it was shown that the conductivity can be significantly increased 
with no changes in the spectral transmittance upon doping with Mo 
\cite{Meng,Yoshida,Yoshida-effmass,Sun}.
Surprisingly, introduction of the transition metal Mo which donates 
two more carriers per substitution compared to Sn, 
does not lead to the expected increase of the optical absorption 
or a decrease of the mobility due to the scattering on the localized Mo d-states.
Moreover, d-states of pre-transition metals (such as Sc and Y) were shown \cite{CdO} 
to affect the band structure of the host TCO material by lowering
the dispersion of the conduction band that, in turn, increases
the effective mass and hence decreases the mobility.
These inconsistencies call for in-depth theoretical analysis.

In this Letter, we present first-principles band structure investigations
of the structural, electronic, optical and magnetic properties of Mo-doped 
In$_2$O$_3$ which reveal that the transition metal dopants can
lead to the transport and optical properties competing with those of 
commercially utilized TCO.
Most importantly, we find that the {\it magnetic interactions} which have 
never been considered to play a role in combining optical transparency with 
electrical conductivity, ensure both high carrier mobility and 
low optical absorption in the visible range.
Compared to ITO, our results show that the magnetic exchange interaction
splits the Mo d-states enabling
(i) smaller increase of the effective mass;
(ii) larger fundamental band gap; (iii) lower short-wavelength optical absorption; 
and (iv) lower plasma frequency and thus lower absorption in the long-wavelength range.
We also demonstrate that strong sensitivity of the d-states of the dopant to 
its oxygen surrounding opens up a possibility to control the transport and optical 
properties with a proper d-element doping.

{\it Approach.} Full-potential linearized augmented plane wave 
method \cite{FLAPW} with the local density approximation is employed \cite{details}
to study the electronic band structure of pure In$_2$O$_3$, ITO and IMO (both 6.25\% doped), 
the latter with and without interstitial oxygen. 
For all systems containing Mo atoms, both non-spin-polarized and 
spin-polarized calculations were performed.
The equilibrium relaxed geometry of all structures investigated 
was determined via the total energy and atomic forces minimization for 
the lattice parameter $a$ and the internal atomic positions. During the optimization, 
all atoms were allowed to move in x, y and z directions. 
Less than 0.5\% increase of the lattice parameter was found upon introduction 
of Mo and the interstitial oxygen, in
agreement with experimental observations \cite{Meng}. The Mo-O distances 
decrease by 4-20\% as compared to the optimized In-O distances in pure In$_2$O$_3$. 

\begin{figure}
\centerline{
\includegraphics[width=8.0cm]{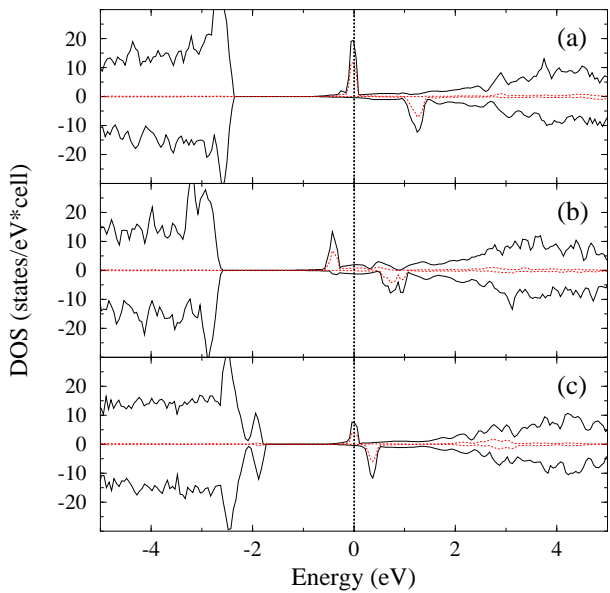}
}
\caption{Spin-resolved total (solid line) and Mo-4d (dashed line) 
density of states for (a) Mo$^{\bullet\bullet\bullet}_{In(1)}$, 
(b) Mo$^{\bullet\bullet\bullet}_{In(2)}$
and (c) [Mo$^{\bullet\bullet\bullet}_{In}$O$^{''}_i$]$^{\bullet}$ structures
(see text for notations). 
}
\label{dos}
\end{figure}

{\it Structural peculiarities.}
In$_2$O$_3$ has the ordered vacancy structure \cite{str} with 8 formula units
where four In atoms out of 16 occupy the centers of the trigonally distorted 
octahedra (denoted as In(1) position), while the rest are located at 
the centers of the tetragonally distorted octahedra (denoted as In(2)).
From the total energy comparison of the 
structures with the Mo atom substituted into In(1) or In(2)
sublattice, we found that 
Mo atoms prefer In(1) positions with the energy difference of 
40 meV that is similar to ITO \cite{Mryasov,exp}.
However, in contrast to the Sn-doped case \cite{Mryasov}, the electronic structures 
of Mo$_{In(1)}$ and Mo$_{In(2)}$ systems are different.
Clearly, the Mo d-states are more sensitive to the oxygen surrounding 
than the spherically-symmetric s-states of Sn atoms: we found that when Mo 
is in the trigonally distorted oxygen octahedron, the three $t_{2g}$ levels 
are degenerate and cross the Fermi level (E$_F$),
while they split by $\sim$0.8 eV into two occupied and one empty level 
in case of the tetragonal distortion of the Mo
oxygen neighbors, cf., Figs. \ref{dos}(a) and \ref{dos}(b).
These findings may help elucidate the unusual observation that the carrier
concentration decreases with temperature (for the systems grown with no
ambient oxygen) which was not previously understood \cite{Yoshida-effmass}.
When the temperature is increased, the second substitutional site, 
i.e., Mo$_{In(2)}$, may become active. In this case, the density of states
at E$_F$ is significantly reduced (cf., Fig. \ref{dos}(a) and \ref{dos}(b)).
Due to the splitting of the d-states, two of the three $t_{2g}$ electrons become
bound and cannot contribute to the charge transport.

{\it Role of interstitial oxygen.}
To investigate the effect of interstitial oxygen on the structural, electronic 
and optical properties of IMO, we calculated the electronic 
band structure and the formation energies \cite{form} of the systems with additional
oxygen located at one of the vacancy positions near the dopant or In atoms.
Using Kr{\"o}ger-Vink notation, the following complexes were considered 
in addition to the Mo$^{\bullet\bullet\bullet}_{In(1)}$ and 
Mo$^{\bullet\bullet\bullet}_{In(2)}$ defects: 
[Mo$^{\bullet\bullet\bullet}_{In(1)}$ + 2In$_{In}$O$^{''}_i$]$^{\bullet}$ and 
[Mo$^{\bullet\bullet\bullet}_{In}$O$^{''}_i$]$^{\bullet}$, 
where the subscript stands for the site position and the superscript stands for 
effective negative ($^{'}$) or positive ($^{\bullet}$) charge. (Note, that 
we omit the site position subscript in the later complex since an interstitial 
oxygen makes the In(1) and In(2) positions equivalent by occupying 
one of the two available vacancy positions.)
First, from a comparison of the formation energies, we found that the addition 
of oxygen results in the energy gain of $\sim$3 eV as compared to the structures 
without an interstitial oxygen, i.e., Mo$^{\bullet\bullet\bullet}_{In}$. 
This finding confirms previous predictions \cite{Meng,Yoshida,Sun} 
that oxygen interstitials are likely to exist in IMO.
We also found that O$_i$ prefers to be located 
near the dopant rather than to be shared by two In atoms, 
forming [2In$_{In}$O$^{''}_i$]$^{\bullet}$ complex,
with the energy difference of 1 eV.

Due to the high probability of the oxygen compensated complexes, one can expect 
a significant reduction of the ionized impurity scattering in IMO. 
Indeed, an increase of the relaxation time was observed 
when ambient oxygen content was increased during the deposition \cite{Yoshida-effmass}.
On the other hand, the interstitial oxygen significantly reduces the 
number of free carriers -- from 3 to 1 per Mo substitution \cite{neutral}. 
Thus, by varying the amount of oxygen one can concurrently control both 
the mobility (through the relaxation time) and free-carrier concentration.
The highest electron concentration can be attained in samples grown 
in 100\% Ar environment \cite{Yoshida-effmass}, i.e., without oxygen,
while the relaxation time should be the shortest in this case and, moreover, 
it should rapidly decrease with Mo concentration since 
Mo$^{\bullet\bullet\bullet}_{In}$ is a strong scattering center.
As the amount of oxygen increases, facilitating the formation of 
[Mo$^{\bullet\bullet\bullet}_{In}$O$^{''}_i$]$^{\bullet}$ complexes, 
the carrier concentration decreases \cite{Yoshida-effmass,Sun}, whereas 
the mobility is improved due to the longer relaxation times.
Thus, both intermediate dopant and oxygen concentrations will provide 
the optimum balance between carrier concentration and mobility resulting in 
the best transport properties of IMO. (Note that as our additional calculations 
showed \cite{w}, introduction of oxygen vacancies which may play an important role 
in samples grown with no oxygen ambient pressure does not affect the conclusions made 
in the current work.)

{\it Role of magnetism.}
Most strikingly, we found that the magnetic configuration is considerably 
lower in energy as compared to the non-magnetic one, and Mo atoms possess a large 
magnetic moment (Table \ref{table}) -- despite the fact that Mo is not magnetic in bulk.
Due to the strong exchange interactions, the Mo d-states located in the vicinity of 
E$_F$ are split by about 1.3 eV and 0.4 eV for Mo$^{\bullet\bullet\bullet}_{In(1)}$ and 
[Mo$^{\bullet\bullet\bullet}_{In}$O$^{''}_i$]$^{\bullet}$, 
respectively, Figs. \ref{dos}(a) and \ref{dos}(c). 
To determine the magnetic coupling between two dopants, we performed
calculations of Mo-doped In$_2$O$_3$ with doubled unit cell.
The total energy difference between ferro- and antiferro-magnetic configurations 
for both Mo$^{\bullet\bullet\bullet}_{In(1)}$ and 
[Mo$^{\bullet\bullet\bullet}_{In}$O$^{''}_i$]$^{\bullet}$ 
is found to be negligible, $\sim$10 meV, suggesting a very weak magnetic coupling 
between the dopants which may result from the screening 
by the free carriers in the system \cite{magn-exp}.

\begin{figure}
\centerline{
\includegraphics[width=4.2cm]{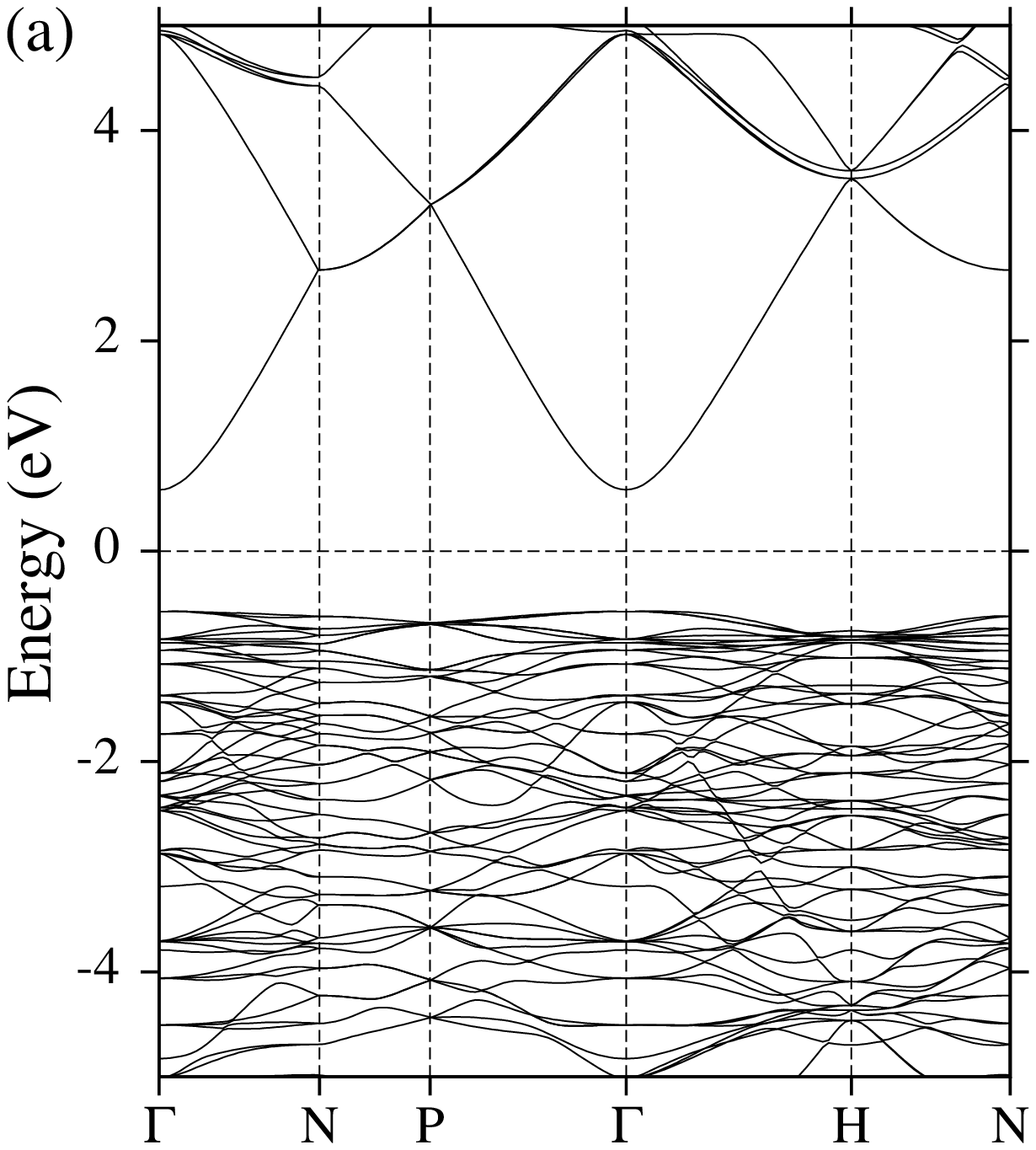}
\includegraphics[width=4.2cm]{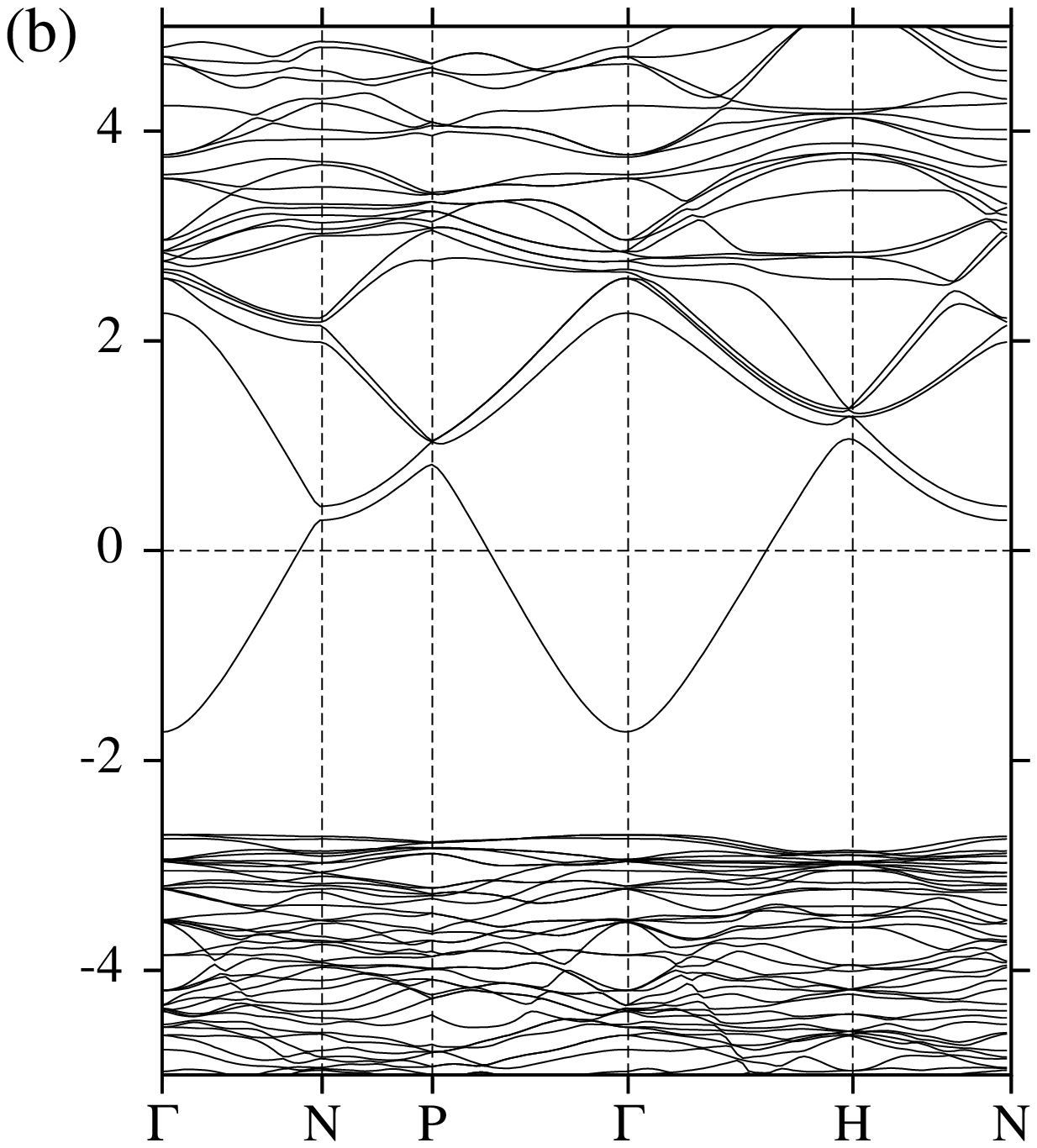}
}
\centerline{
\includegraphics[width=4.2cm]{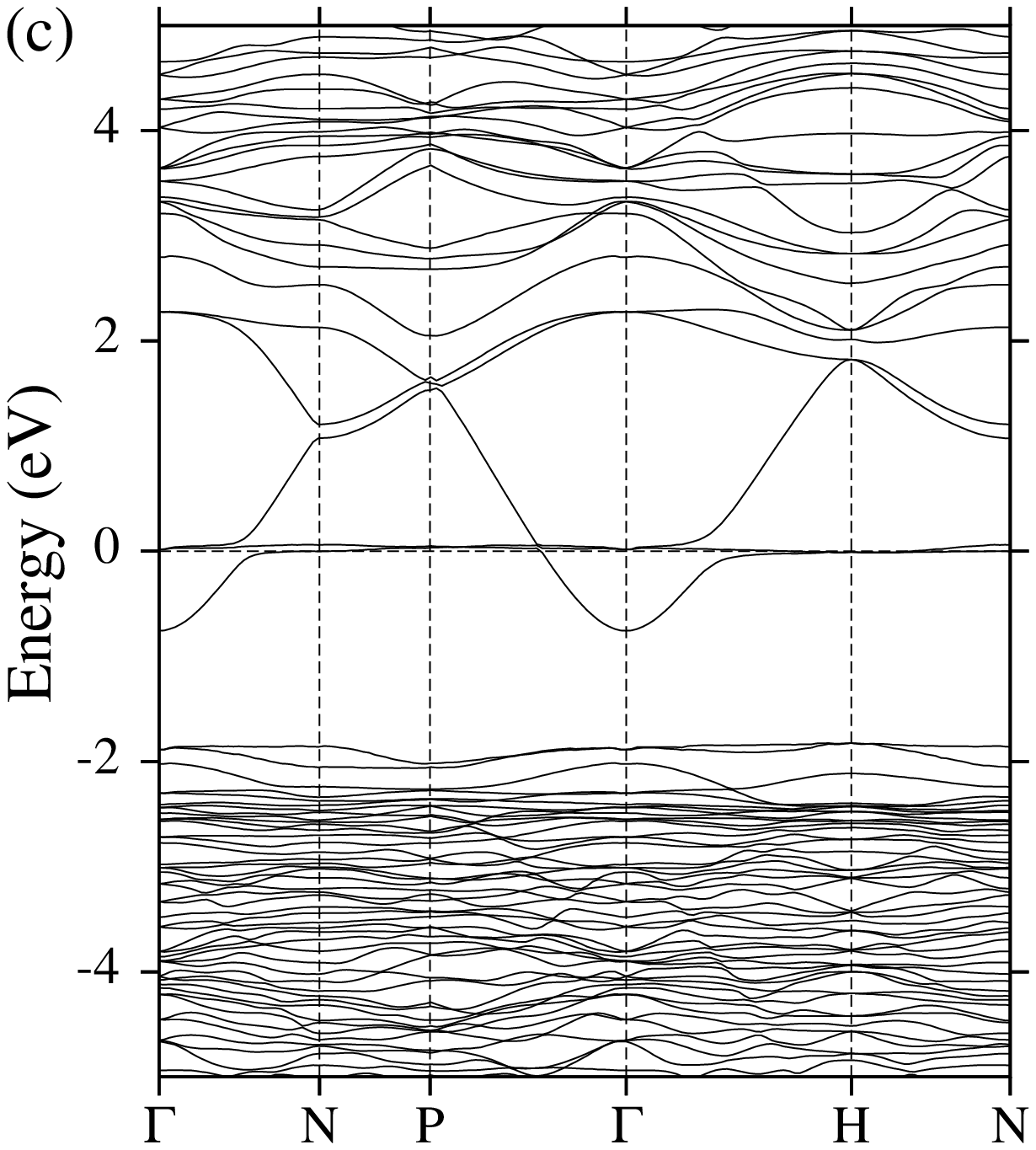}
\includegraphics[width=4.2cm]{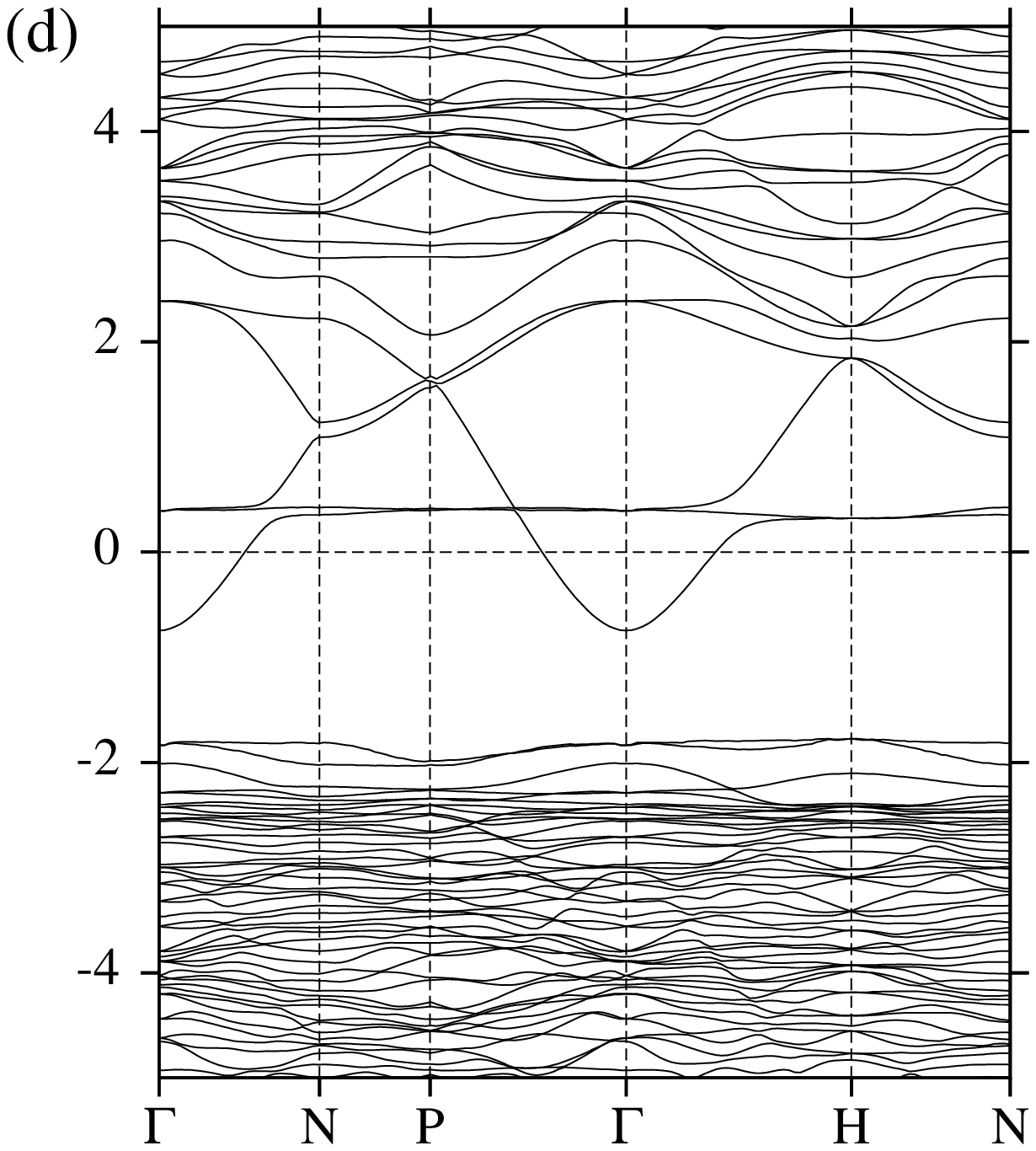}
}
\caption{
Band structure for (a) pure In$_2$O$_3$, (b) Sn$^{\bullet}_{In(1)}$
structure, and (c) the majority spin and (d) the minority spin channels 
of [Mo$^{\bullet\bullet\bullet}_{In}$O$^{''}_i$]$^{\bullet}$ structure. 
}
\label{bands}
\end{figure}

It is important to note that the conductivity in IMO is due to the delocalized
In s-states which form the highly disperse free-electron-like conduction band 
(cf., Fig. \ref{bands}), while the Mo d-states are resonant states.
The free carriers in the system flow in a background of the Mo defects
which serve as strong scattering centers.
Because of the exchange splitting of the Mo d-states, the carriers of one spin will 
be affected by only a half of the scattering centers, i.e., only by the Mo 
d-states of the same spin. Therefore, the concentration of the Mo scattering
centers is effectively lowered by half compared to the Mo doping level.
In other words, the lack of long-range magnetic order should lead to the formation of two 
interpenetrating networks transporting efficiently the carriers of opposite spin.
Moreover, Mo population of the In(2) positions at elevated temperatures
should enable the high conductivity in {\it both} spin channels since both the 
majority and minority Mo d-states are pushed away from E$_F$, Fig. \ref{dos}(b).

Because the interstitial oxygen significantly suppresses the magnetic interactions, 
Table 1, the optimum transport properties can be achieved at intermediate 
oxygen concentration. As discussed above, the interstitial oxygen increases 
the relaxation time by reducing the average charge of the substitutional complexes
(note the decrease of the density of states peak at E$_F$ upon introduction of 
O$_i$ which binds two electrons and forms a new band at the top of the valence band, 
cf., Figs. \ref{dos}(a) and \ref{dos}(c)).
On the other hand, the magnetic exchange interactions should be strong enough to split 
the transition metal d-states in order to provide conductivity in one or both spin channels.

{\it Burstein-Moss (BM) shift.}
To illustrate how the transition metal doping alters the electronic band structure of 
the host material and to compare the changes with the traditional Sn-doped case, 
we also calculated the band structures of pure and 6.25\% Sn-doped In$_2$O$_3$.
The resulting band structure plots are presented in Fig. \ref{bands}. In addition,
we compare the calculated fundamental band gaps E$_g$(0), the Fermi wave vectors k$_F$ 
and the plasma frequencies $\omega_p$ \cite{plasma} in Table \ref{table}.
Most significantly, we found that the BM shift (i.e., the displacement of 
E$_F$ above the conduction band minimum upon doping) is less pronounced 
in the IMO case -- despite the fact that, at the same doping level,
Mo$^{6+}$ donates two extra carriers as compared to Sn$^{4+}$. 
This can be clearly seen from a comparison of k$_F$  
for the Mo$^{\bullet\bullet\bullet}_{In(1)}$ and Sn$^{\bullet}_{In(1)}$ structures.
(Note, that unlike IMO, the doping with Sn results in k$_F$ and $\omega_p$ to be 
similar to those found from the rigid-band shift in pure In$_2$O$_3$.)
Such a low sensitivity to doping appears from the resonant Mo d-states
located at E$_F$ that facilitates the d-band filling and thus hinders further 
displacement of E$_F$ deep into the conduction band.

{\it Effective mass and optical properties.}
Smaller BM shift in IMO arising from the electron localization on the Mo d-orbitals
(pinning) leads to the following advantageous features to be compared to those of ITO:

(i) Smaller increase in the effective mass is expected upon Mo doping.
In addition, the Mo d-states do not hybridize with the s-states of indium which form
the highly disperse 
conduction band. Therefore,
in contrast to ITO with the strong hybridization between In and Sn s-states \cite{Mryasov},
the resonant Mo d-states at E$_F$ do not affect the dispersion of the conduction band
and hence the effective mass remains similar to the one of pure indium oxide. 
This is borne out in experimental observations \cite{Yoshida-effmass} showing
that the effective mass does not vary with doping (up to 12 \% of Mo)
and/or carrier concentration.

(ii) Smaller BM shift does not lead to the appearance of the intense interband transitions 
from the valence band in the visible range since the optical band gap 
in pure indium oxide is large enough, namely 3.6 eV \cite{Hamberg}.
Furthermore, in contrast to ITO where E$_g$(0) narrowing 
has been demonstrated both experimentally \cite{Hamberg} 
and theoretically \cite{Mryasov}, 
doping with Mo shows an opposite (beneficial) trend:
we found that E$_g$(0) increases upon introduction of Mo
by 19 \% and 9 \% for Mo$^{\bullet\bullet\bullet}_{In(1)}$ and
[Mo$^{\bullet\bullet\bullet}_{In}$O$^{''}_i$]$^{\bullet}$, respectively, 
while in the case of Sn it decreases by 16 \% -- as compared 
to pure In$_2$O$_3$ (Table \ref{table}).
We believe that the widening of E$_g$(0) appears from 
the anisotropic d-orbitals of Mo which, in contrast to spherical s-orbitals of Sn,
rotate the p-orbitals of the oxygen neighbors. As a result, the overlap of these 
p-orbitals with the In s-states increases, leading to the widening of E$_g$(0).

(iii) Larger (in energy) optical transitions from the partially occupied band, i.e.
from E$_F$, up into the conduction band (cf., Figs. \ref{bands}(b) and 
\ref{bands}(c,d)) along with the fact that transitions from d- to s-states 
are forbidden ensure lower short-wavelength optical absorption \cite{2gap}.

(iv) The calculated $\omega_p$ in IMO is below the visible range and 
significantly smaller than that of ITO (Table \ref{table}).
This finding suggests a possibility to introduce larger carrier concentrations
without sacrificing the optical transmittance in the long wavelength range.

Finally, 
we found that the interstitial oxygen 
results in a decrease of both E$_g$(0) and BM shift 
(Table \ref{table}) -- in agreement with the observed decrease of 
the optical band gap with ambient oxygen content \cite{Sun}. Nevertheless, the optical 
band gap remains well above the visible range. Moreover, our calculations suggest 
that the overall transmittance should increase with oxygen concentration
due to the increased energy of the interband transitions from E$_F$
(smaller BM shift) and the lower plasma frequency.

In summary, using IMO as an example, we demonstrate that the transition
metal dopants can be highly beneficial in providing the transport and optical properties 
which compete with those of commercial TCO materials. Our detailed first-principles 
investigations reveal the intricate role of magnetic interactions in determining both
the electrical conductivity and the BM shift which in turn governs 
optical absorption in the visible range.
Furthermore, additional 
calculations of In$_2$O$_3$ doped with W and 3d-elements \cite{w} 
show that similar behavior can be observed in other systems 
leading to a family of efficient transparent conductors mediated 
by magnetic interactions. Recent experiments \cite{Tate} support this conclusion.
We expect that a variety of the appealing features in the electronic band structure
arising from the electronic configuration of a proper transition metal dopant 
and a strong sensitivity of the d-states to the oxygen surrounding will 
allow control over the transport and optical properties of these advanced
TCO materials.

Author acknowledges J. Tate for stimulating discussions.
The work is supported by the University of Missouri Research Board.

\begin{table}
\begin{tabular}{c|ccc|ccc|c} \hline
Complex & $\Delta$E$_{tot}$ & M & E$_g$(0) & k$_F^{\Sigma}$ & k$_F^{\Lambda}$ & k$_F^{\Delta}$ & $\omega_p$ \\ \hline

Mo$^{\bullet\bullet\bullet}_{In(1)}$ & 563 & 1.85 & 1.38 & 0.152 & 0.148 & 0.157 & 1.63 \\

Mo$^{\bullet\bullet\bullet}_{In(2)}$ & 370 & 1.32 & 1.18 & 0.194 & 0.187 & 0.201 & 2.05 \\

[Mo$^{\bullet\bullet\bullet}_{In}$O$^{''}_i$]$^{\bullet}$ & 32 & 0.50 & 1.26 & 0.125 & 0.123 & 0.132 & 1.27 \\

[Mo$^{\bullet\bullet\bullet}_{In(2)}$+2InO$^{''}_i$]$^{\bullet}$ & 112 & 0.61 & 1.06 & 0.120 & 0.118 & 0.123 & 1.24 \\ \hline

Sn$^{\bullet}_{In(1)}$ & & & 0.98 & 0.201 & 0.203 & 0.205 & 2.29 \\ 

In$_2$O$_3$+e$^{'}$ & & & 1.16 & 0.206 & 0.204 & 0.213 & 2.38 \\ \hline

\end{tabular}
\caption{The
total energy differences $\Delta$E$_{tot}$ between non-magnetic and ferromagnetic 
configuration, in meV; the magnetic moments, M, on the Mo atoms, in $\mu_B$;
the fundamental band gap values E$_g$(0), in eV; the Fermi wave vectors k$_F$, in atomic
units, along the [110]($\Sigma$), [111]($\Lambda$) and [010]($\Delta$) directions; 
and the plasma frequency $\omega_p$, in eV,
for the different substitutional complexes with 6.25\% Mo doping level.
Calculated values for pure (rigid-band model) and 6.25\% Sn-doped In$_2$O$_3$ are also given.}
\label{table}
\end{table}



\begin{thebibliography}{99}
\bibitem{form}
The formation energy is calculated as 
E$_f$ = E$_{In_2O_3:Mo}$ -- E$_{In_2O_3}$ -- E$_{In}$ + E$_{Mo}$,
where E$_{In_2O_3:Mo}$, E$_{In_2O_3}$, E$_{In}$, and E$_{Mo}$ are
the total energy of the Mo-doped and the corresponding undoped indium oxide, 
and bulk bct
In and bcc
Mo, respectively. 
For the structures with interstitial oxygen, the E$_f$ was calculated
in two ways: via the total energy of indium oxide with an additional oxygen, 
i.e., E$_f$ = E$_{In_2O_3:MoO_i}$ -- E$_{In_2O_3:O_i}$ -- E$_{In}$ + E$_{Mo}$
and the total energy of the oxygen molecule O$_2$, i.e.,
E$_f$ = E$_{In_2O_3:MoO_i}$ -- E$_{In_2O_3}$ -- E$_{In}$ + E$_{Mo}$ + E$_{O_2}$/2.
\bibitem{neutral}
This is in contrast to ITO, where interstitial oxygen
observed experimentally \cite{Mason} results in 
the formation of neutral defect complexes, [2Sn$^{\bullet}_{In}$O$^{''}_i$]$^x$.
In IMO, neutral complexes may arise from
[2Mo$^{\bullet\bullet\bullet}_{In}$3O$^{''}_i$]$^x$. Another possibility 
is that Mo may also exist in the valence state of 4+ which may lead to
the formation of [2Mo$^{\bullet}_{In}$O$^{''}_i$]$^x$ complexes. 
Studies of the neutral complexes are beyond the scope of this work.
\bibitem{magn-exp}
Weak coupling between the dopants makes it hard to detect the magnetic properties 
experimentally.
\bibitem{plasma}
The plasma frequency was estimated using the Eq. (1) in Ref. \cite{Mryasov} 
via the calculated Fermi wave vectors and Fermi group velocities.
\bibitem{2gap}
Our results show that the second 
(hybridization) gap which was argued to be one of the reasons of low optical 
absorption in ITO \cite{Mryasov} closes upon structural relaxation.
\end{thebibliography}
\end{document}